# Automated Supervised Identification of Thunderstorm Ground Enhancements (TGEs)

Davit Aslanyan
Alikhanyan National Laboratory (Yerevan Physics Institute), Yerevan 0036, Armenia
E-mail: davitaslanian@gmail.com

**Abstract**

Thunderstorm Ground Enhancements (TGEs) are bursts of high-energy particle fluxes detected at Earth's surface, linked to the Relativistic Runaway Electron Avalanche (RREA) mechanism within thunderclouds. Accurate detection of TGEs is vital for advancing atmospheric physics and radiation safety, but event selection methods heavily rely on expert-defined thresholds. In this study, we use an automated supervised classification approach on a newly curated dataset of 2024 events from the Aragats Space Environment Center (ASEC). By combining a Tabular Prior-data Fitted Network (TabPFN) with SHAP-based interpretability, we attain 94.79% classification accuracy with 96% precision for TGEs. The analysis reveals data-driven thresholds for particle flux increases and environmental parameters that closely match the empirically established criteria used over the last 15 years. Our results demonstrate that modest but concurrent increases across multiple particle detectors, along with strong near-surface electric fields, are reliable indicators of TGEs. The framework we propose offers a scalable method for automated, interpretable TGE detection, with potential uses in real-time radiation hazard monitoring and multi-site atmospheric research.

**Introduction**

Charge separation in thunderclouds, driven by updrafts of warm air and interactions among hydrometeors, creates oppositely directed dipoles within the cloud. The oppositely directed atmospheric electric field (AEF) in the upper and lower dipoles accelerates free electrons toward open space and the Earth's surface. Free electrons are abundant in the atmosphere due to extensive air showers (EASs). Electric fields generated by strong thunderstorms transfer energy to these electrons, accelerating them and, under certain conditions, leading to electron-photon avalanches. These avalanches propagate through large volumes of the atmosphere, covering wide areas when they reach the Earth's surface, significantly increasing natural gamma radiation (NGR), which affects radiation safety and climate. These enhancements can last from seconds to tens of minutes (Chilingarian et al., 2020).
The ionized channels formed by relativistic electrons create pathways for lightning leaders to move toward the ground (Chilingarian et al., 2017). A key physical process behind these atmospheric particle flux increases is the Relativistic Runaway Electron Avalanche (RREA) process, introduced by Gurevich et al. (1992). RREAs are crucial for understanding a range of high-energy atmospheric phenomena, including Thunderstorm Ground Enhancements (TGEs, Chilingarian et al., 2010, 2011), gamma-ray glows (Marisaldi et al., 2024), and both upward and downward terrestrial gamma-ray flashes (TGFs, Fishman et al., 1994).

Recognizing the shared physical origin of brief, microsecond-scale TGFs, minute-long gamma glows, and surface-level TGEs is essential for understanding high-energy atmospheric physics and marks a step towards accepting RREA and EAS as a universal physical process that is responsible for the enhanced particle fluxes in the lower and upper atmosphere (Chilingarian et al., 2022a, 2023, 2024a).

While TGFs consist of microsecond bursts of gamma radiation originating in equatorial thunderstorms and observed from orbiting gamma-ray observatories positioned 400 to 700 kilometers above the source, TGEs manifest as intense, prolonged particle fluxes detectable at ground level that originate from accelerating electric fields located directly above particle detectors (often within 25-100 m), enabling detailed measurement of electron and gamma-ray energy spectra and characterization of thundercloud charge structures.

Gamma glows represent gamma-ray radiation from RREA within a thunderstorm's upper dipole and are typically detected at higher altitudes by balloon or aircraft-based instruments. These emissions last from tens of seconds to several minutes, often ending with lightning discharges. Gamma-ray emissions observed at Earth's surface are sometimes also referred to as gamma glows due to their exclusive gamma-ray content, which indicates the altitude of the thunderclouds and the absorption of lower-energy particles.

Given the variety of physical mechanisms that influence atmospheric particle fluxes, accurately identifying specific processes is essential. Since the initial observations at Aragats in 2009, over 1,000 TGE events have been recorded at mountain observatories worldwide, including locations in Eastern Europe, Japan, Russia, Germany, and Armenia (Chilingarian et al., 2025 and references therein). Recent observations extend these findings to Mt. Hermon in Israel (Mauda et al., 2025) and to sites in Finland (Leppänen et al., 2025) and Slovakia (Kísvardai et al., 2025). Along with numerical simulations, these observations offer detailed insights into the RREA mechanism and related cloud-charge distributions within the lower atmosphere.

In this work, we expand the study of TGEs by applying a supervised classification model to a newly curated dataset of 2024 events. Using the Tabular Prior-data Fitted Network (TabPFN) model and SHapley Additive exPlanations (SHAP)-based interpretability analysis, we derive data-driven thresholds for key variables and demonstrate their consistency with manual criteria developed over 15 years of observational experience.

## 2. Methods

The Aragats Space Environment Center (ASEC), located at the Aragats high-altitude research station of the Cosmic Ray Division, is situated at 3200 meters on Mt. Aragats in Armenia. It experiences frequent and intense thunderstorms during spring and summer, often causing thunderclouds to descend below 100 meters above the detectors. Low-altitude thunderclouds create a significant electric field gradient between the main negative (MN) charge layer and its mirror image on Earth's surface, which promotes the RREA process registered at the Earth's surface as TGE. This gives ASEC a uniquely advantageous position for detecting particle bursts.

### 2.1. Detectors

Throughout its 80 years of continuous operation, ASEC has integrated various instruments for detecting different species of cosmic rays, electric field disturbances, and meteorological conditions. In this study, we analyze data from the STAND 1cm, STAND 3cm, SEVAN particle detectors, BOLTEK EFM-100 electric field mills, and automatic weather stations from DAVIS Instruments. The detailed descriptions of these detectors can be found in Chilingarian et al. (2024b). The STAND 1cm detector consists of a three-layer assembly of 1 cm-thick molded plastic scintillators with a 1-square-meter sensitive area stacked vertically, along with a 3 cm-thick scintillator positioned nearby (Figure 1).

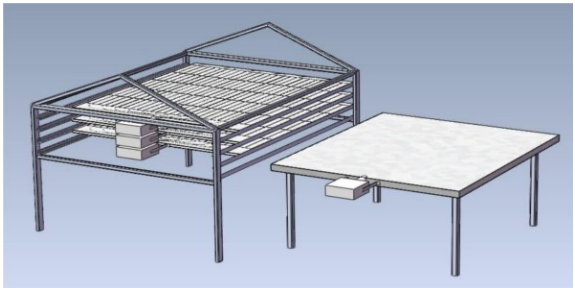

**Figure 1. The stacked STAND 1cm detector consisting of three 1 cm thick, 1 m² area scintillators and a stand-alone 3cm thick scintillator of the same area.**

Light from the scintillator is re-emitted in the long-wavelength range by optical spectrum-shifter fibers and then directed to the FEU-115M type photomultiplier (PMT). The peak luminescence occurs at 420 nm, with a luminescence decay time of approximately 2.3 ns. The high voltage and discrimination threshold of the photomultiplier are adjusted so that the upper scintillator of the 1 cm STAND detector has an energy threshold of about 0.8 MeV. The detector is integrated into a fast data synchronization system capable of capturing time series at a 50-ms sampling rate, precisely synchronized with atmospheric discharges at nanosecond accuracy. The 3 cm STAND detector consists of a four-layer assembly of 3 cm-thick plastic scintillators with a 1-square-meter sensitive area stacked vertically (Figure 2). The detector electronics operate similarly to the 1 cm STAND detectors, and by registering coincidences, can differentiate electrons with energies of 5, 20, 30, and 40 MeV.

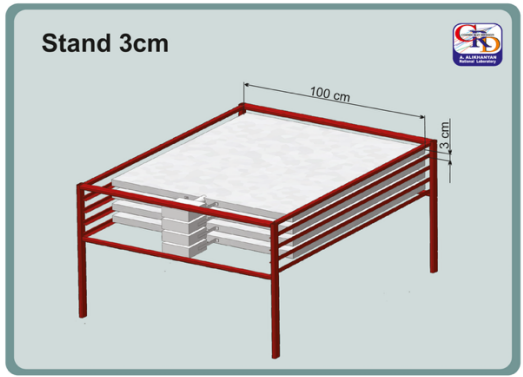

Figure 2. The Stand 3cm detector consisting of four stacked 3 cm thick scintillators with an area of 1 m².

SEVAN is a hybrid particle detector that measures gamma rays, neutrons, muons, positrons, and electrons. It features a three-layer system made up of plastic scintillator slabs, lead absorbers, light guides, and PMTs. SEVAN consists of two identical assemblies of plastic scintillator slabs, each measuring 100 cm x 100 cm x 5 cm. Between these assemblies, a thick scintillator stack measuring 50 cm x 50 cm x 20 cm is positioned (

). The upper scintillator has a threshold of about 7 MeV.

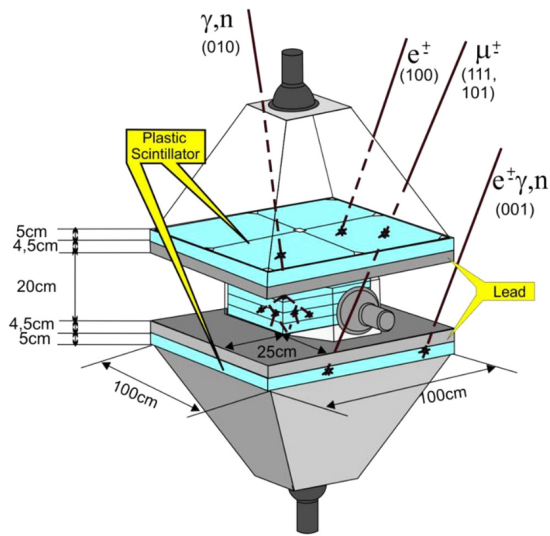

Figure 3. Assembly of SEVAN Detecto

Information about the Boltek EFM-100 electric field mill and Davis Vantage Pro2 weather station (see one of the locations of these detectors on the roof of "Cuckoo's Nest" lab in Figure 4) can be found in Chilingarian et al. (2024c).

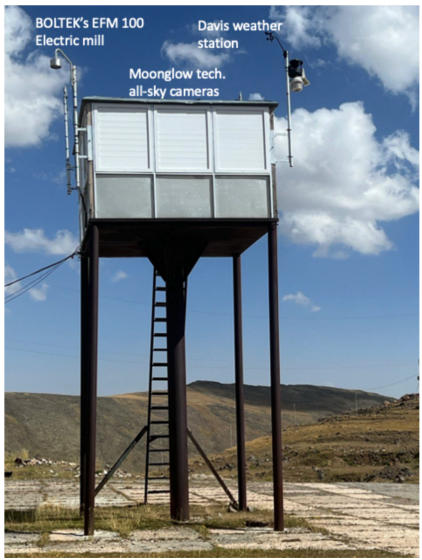

**Figure 4. Electric field mill EFM 100 from BOLTEK, DAVIS weather station, and all-sky cameras installed on the roof of a small "Cuckoo's Nest" lab.**

## 2.2. TGE Event definition

TGEs are classified using a detailed, multi-parameter method based on raw data collected over a 15-year span. This approach combines multi-detector monitoring of various cosmic ray species, near-surface electric fields (NSEF), and environmental data to confirm that each recorded event is related to thunderstorm-induced particle acceleration rather than unrelated background phenomena or equipment failures (Chilingarian et al., 2023).

A candidate TGE event is confirmed only when independent particle detectors record simultaneous and statistically significant increases in count rates. Specifically, the procedure requires that at least three detectors, particularly the SEVAN, STAND1, and STAND3 scintillators, observe a flux rise exceeding three standard deviations (3σ) above pre-storm mean values. Including detectors with higher energy thresholds, such as STAND3 and SEVAN, ensures that the detected increase accurately indicates the presence of RREAs and is not merely caused by lower-energy radon progeny radiation.

Along with the particle flux criteria, each TGE must also happen during a sufficiently strong atmospheric electric field. To be accepted, the absolute value of the NSEF during the event must surpass 5 kV/m. This method allows classifying TGEs based on their strength and duration into two categories: "large" TGEs and "small" TGEs.

Small TGEs generally show only modest increases and limited high-energy electron counts. These less intense events are more common in summer, when higher temperatures cause higher cloud bases that reduce the particle flux reaching the Earth's surface. In Figure 5, we display a summer TGE event that occurred on August 2, 2024. The count rate increases slightly, reaching the 3σ threshold. The NSEF was quite strong, at -20 kV/m; however, the cloud base was very high, above 600 meters. Therefore, possibly, high in the atmosphere, the RREA was very intense, but most particles were absorbed in the thick atmosphere, and only a few gamma rays reached the detector. In any "border" situation for small TGEs, it is important to consider environmental factors and the season.

Large TGEs usually show more than 20% increases in particle detector counts above their normal levels across all detectors considered. These significant events mainly happen in Spring and Fall, when low cloud bases and short free passage distances allow near-surface electron accelerators to operate with exceptional stability (Chilingarian et al., 2024c).

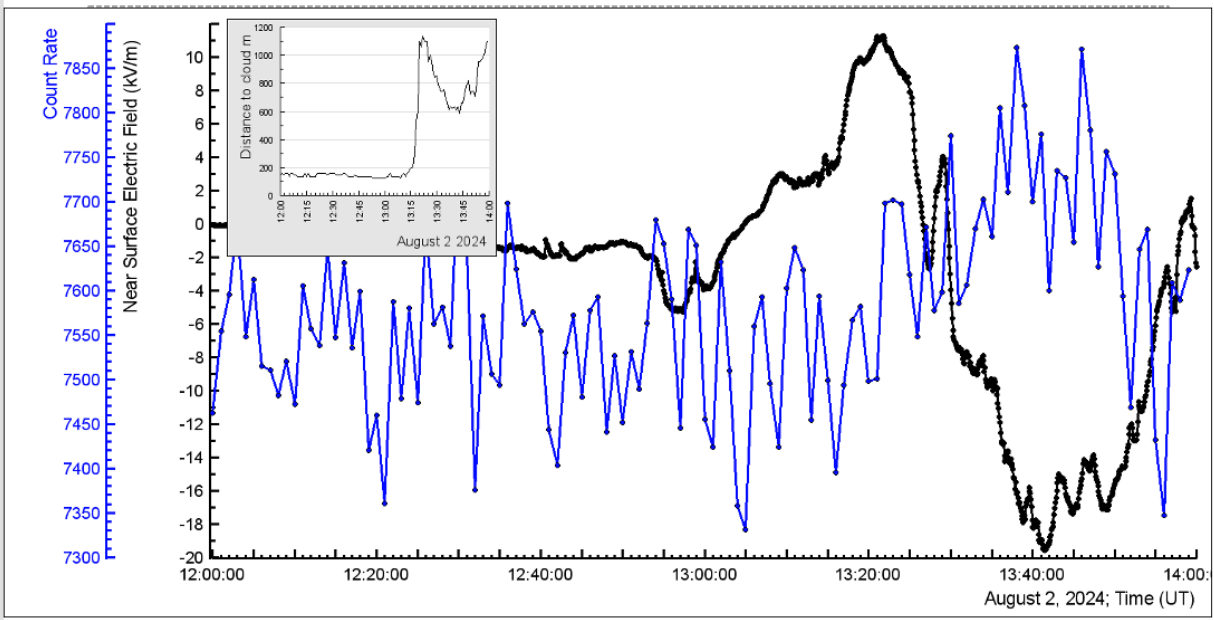

**Figure 5. Small Summer TGE registered by STAND 3cm detector (blue curve); disturbances of electric field are shown in black. In the inset – distance to the cloud base.**

An example of a very intense TGE is shown in Figure 6. To highlight the dynamics of count rate change, we present a 1-second time series. The network of STAND 1cm detectors allowed measurement of particle fluxes with 50 ms sampling. On October 2, 2024 particle detectors on Aragats recorded a double-stage TGE with a sharp increase in particle flux. A one-second measurement from a 1 cm thick, 1 m² outdoor scintillator on the roof of the GAMMA experiment's calorimeter shows nearly a tenfold rise (900%, 120 σ, the second peak). The black curve displays the NSEF, which is strongly negative during TGE. A broad peak began at 00:41:40, peaked at 00:42:35, then declined at 00:43:15 before immediately rising again until a cloud-to-ground (-CG) lightning flash abruptly ended the TGE at 00:43:36. The RREA electron flux in the cloud was sufficient to create an ionization channel in the lower atmosphere, providing a path for the lightning leader (see Chilingarian et al., 2017). This was the first and only TGE of 2024 with a large electron content. The increase in low-energy gamma-ray and electron flux detected by the STAND3 plastic scintillator (see Chilingarian and Hovsepyan, 2023) reached 225% (125 σ). The 20-cm-thick, 0.25-m² area spectrometric SEVAN light scintillator (Chilingarian et al., 2024d) detected electrons with energies above 10 MeV.

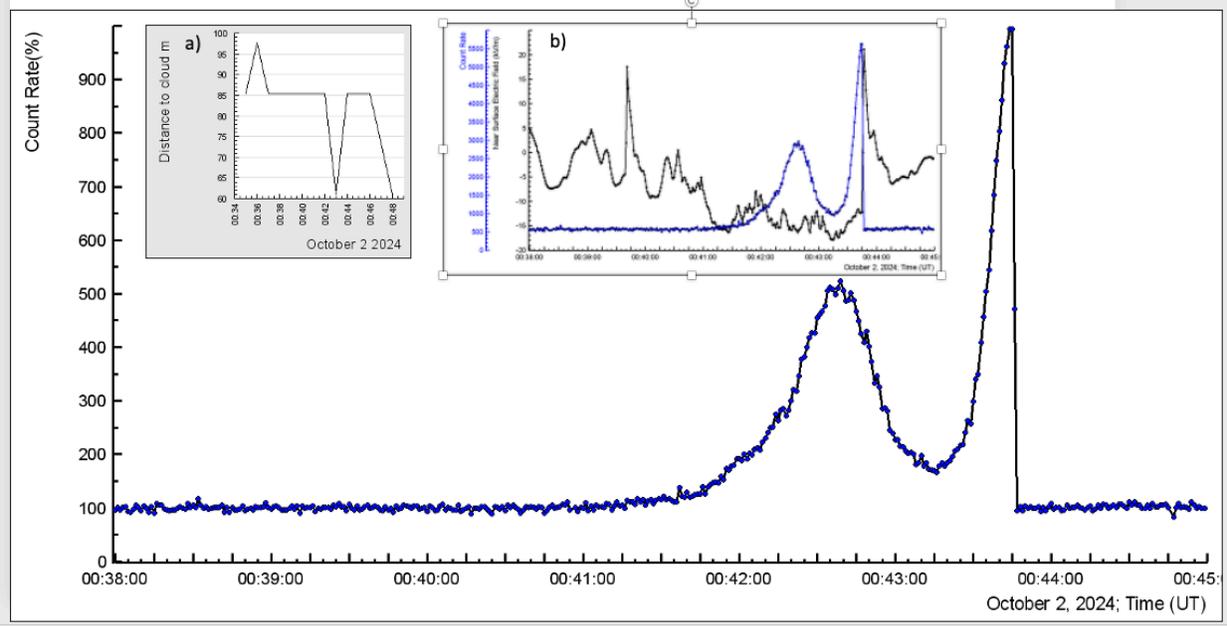

**Figure 6. 1-second time series of a large TGE event observed by the upper scintillator of the STAND 1cm detector. In the inset, a) shows the distance to the cloud base, and b) displays disturbances in the electric field.**

Throughout previous studies, the application of this strict selection procedure has resulted in a well-curated catalog of verified TGEs, none of which have required subsequent retraction (Chilingarian et al., 2022b, 2024a, 2024b). In this study, we use the same methodology to build a dataset of TGE events observed in 2024, which serves as the ground-truth reference for training and evaluating our new automated analysis methods.

## 2.3. TGEs observed in 2024

The 2024 core variables include data from the STAND 1cm, STAND 3cm, and SEVAN detectors, recorded as both percentage increases and statistical significance compared to pre-storm background levels. The detectors measure particle flux via their upper scintillators, which count all passing particles, while coincidence channels (100 for STAND 1cm and SEVAN; 1000 for STAND 3cm) restrict counts to particles that stop in the upper scintillator, thereby selecting low-energy particles.

Measurements of NSEF are included along with a new constructed variable, ΔE, which is defined as the difference between the highest and lowest NSEF values within the identified event window. ΔE thus measures the dynamic variability of the electric field, offering insights into the mechanisms of particle enhancements. All meteorological parameters (temperature, humidity, cloud base height, solar radiation, atmospheric pressure, wind speed, and rain rate) were recorded to ensure correlation studies across variables.

To prevent overcounting of neighboring TGEs, only enhancements that significantly decrease after reaching the peak, before the second subsequent rise, were regarded as distinct TGEs, with at least a ten-minute gap between events. Redundancy within the ASEC network, which includes three electric field mills and two weather stations, minimized the effects of instrument outages; any gaps were filled by redundant sensors.

The dataset also includes training examples with TGE: control samples recorded under fair-weather conditions and during thunderstorms that did not meet TGE confirmation criteria. These entries (TGE = 0) provide the necessary contrast to confirmed events (TGE = 1), ensuring the model learns meaningful distinctions rather than defaulting to single-class prediction. This balanced design underpins the robustness of subsequent classification experiments.

## 2.4. Artificial Intelligence Model and Threshold Determination

This section describes the statistical modeling used to compare the model-derived TGE classification rule with the empirical decision rules for TGE selection from our previous work. In addition, we observe during particle count bursts strong correlations between TGEs and atmospheric parameters. We used the TabPFN, a recently developed artificial intelligence (AI) model specifically designed for small to medium-sized datasets, usually containing up to 10,000 samples (Hollmann et al., 2025). Unlike traditional machine learning methods, which require model selection, feature preprocessing, and hyperparameter tuning, TabPFN is a prior-data fitted model that is trained once and applied directly to new tasks.

TabPFN is based on a transformer architecture, originally developed for sequence modeling in natural language processing, but adapted to the tabular domain. Its training procedure involves exposure to a vast corpus of synthetically generated classification tasks, from which it learns a meta-distribution over tabular problem. This allows the model to approximate a Bayesian posterior predictive distribution over class labels conditioned on input features, effectively functioning as an amortized Bayesian inference engine for tabular data.

In practice, this means that TabPFN can provide out-of-the-box high-quality predictions without iterative retraining, making it particularly well-suited to scenarios in physics where datasets may be limited in size, costly to generate, or not amenable to extensive hyperparameter tuning. By encoding statistical relationships between features and outcomes during pretraining, TabPFN bypasses the sequential model-building process characteristic of ensemble methods such as gradient boosting.

However, a limitation is that TabPFN, like many deep learning models, is intrinsically non-interpretable. Its decision-making relies on distributed representations within the transformer layers, which makes it non-trivial to directly attribute predictions to specific features. For applications where interpretability is essential - for instance, to uncover underlying physical mechanisms - TabPFN must therefore be complemented with post-hoc explainability techniques or contrasted with more transparent baseline models.

Unlike classical regression models, which provide coefficients reflecting the influence of each predictor, "black box" models require additional methods to understand the contributions of each feature. To address this, we utilized SHAP (Lundberg & Lee, 2017; Lundberg et al., 2018), a framework that quantifies each feature's effect on individual predictions, allowing comparison and ranking of variables according to their relative importance. The SHAP values are grounded in game theory and are consistent: as a predictor's effect on the model output increases, its SHAP value rises, and results are insensitive to feature scaling. For each observation, SHAP values across all features sum to zero, reflecting a zero-sum framework. Summary plots provide a global view of feature importance and directional influence. The horizontal axis indicates the SHAP value (impact on the predicted outcome), while the vertical axis ranks features by overall importance. Individual observations are represented by color-coded points reflecting feature magnitude and mean absolute SHAP values are displayed to quantify relative importance within the model.

The model was trained to classify each observation in the 2024 dataset as either TGE or non-TGE using a leave-one-out cross-validation (LOOCV) strategy. In LOOCV approach, each observation is held out as the test set while the model is trained on all remaining observations, ensuring that every observation is used once for validation. This method reduces variance in performance estimates compared to a single 70/30 or 80/20 training/testing split and is especially valuable for datasets of modest size where retaining as much training data as possible improves stability in learned thresholds.

The input features were selected to optimize the signal-to-noise ratio, including percentage enhancements of peak fluxes from STAND1, STAND3, and SEVAN upper scintillators, as well as environmental parameters such as near-surface temperature and the absolute value of the NSEF. Coincidence channels, though physically relevant, were reserved for expert review due to their higher background fluctuations.

To further explore the relationship between each variable and its SHAP value, we plotted SHAP values against raw feature values. These scatterplots can be noisy, complicating threshold determination. To address this, we applied a locally weighted scatterplot smoothing (LOWESS) method (Cleveland, 1979; Cleveland and Devlin, 1988). LOWESS fits a smooth curve through the data without assuming a global functional form, capturing non-linear trends. The point where the smoothed curve crosses zero indicates the feature value at which its influence on TGE prediction changes sign, providing a practical, data-driven threshold for decision-making.

## 3. Results

The final classifier achieved an overall accuracy of 94.79% on the test set, with 96% precision for the TGE class and a recall of 97%, demonstrating strong generalization performance. The confusion matrix showed two false negatives (True label 1, Predicted label 0) and three false positives (True label 0, Predicted label 1) for TGEs (Figure 7).

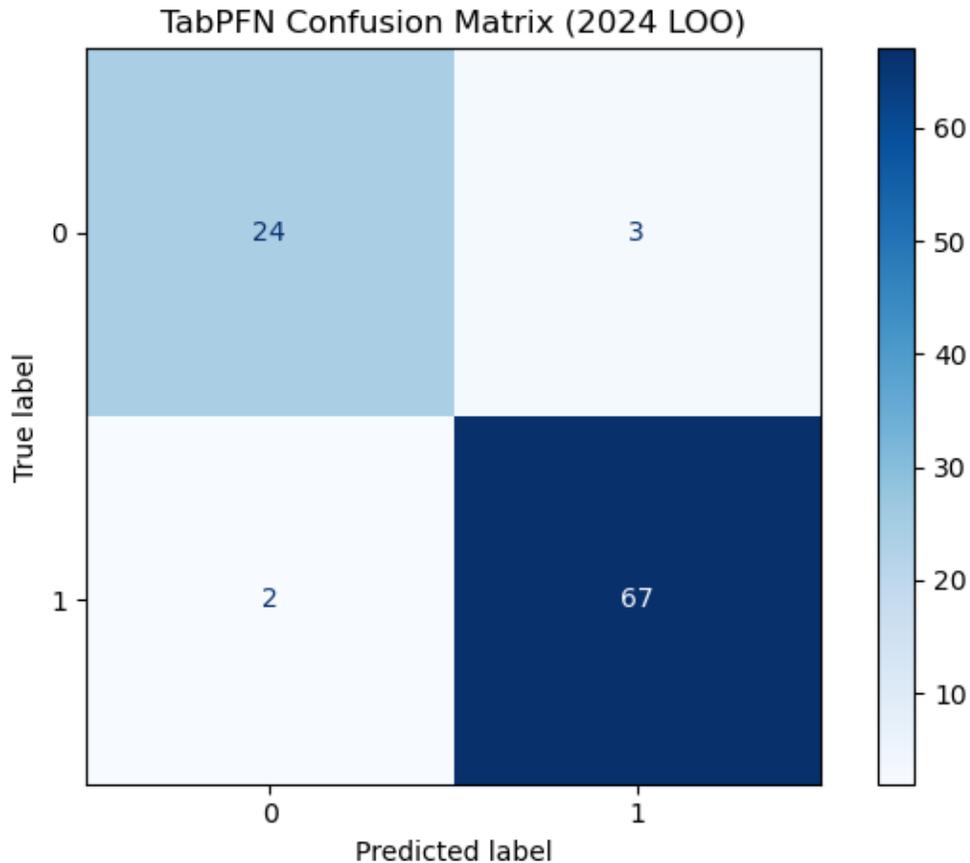

**Figure 7. Confusion matrix for the TabPFN classifier on the held-out test set.**

Figure 8 presents a SHAP "beeswarm" summary plot, where each point corresponds to one observation's SHAP value for a given feature. The horizontal spread reflects the magnitude of each feature's contribution to the model output, while the color scale (cool to warm) shows the raw value of the feature. Features are ordered top to bottom by their mean absolute SHAP values, indicating their overall importance in the classification task. The top three predictors are the percentage enhancements of STAND 3cm and SEVAN upper scintillators (denoted by STAND3 (%) and SEVAN (%) in Figure 8), followed by the absolute value of the near-surface electric field (denoted in Figure 8 by |NSEF|). Contributions from the cloud base height and the outside temperature are smaller but still non-negligible. Overall, increases in particle detector enhancements (represented by warm-colored points on the right) consistently result in positive

SHAP values, indicating that these features tend to push the model's prediction toward the TGE class.

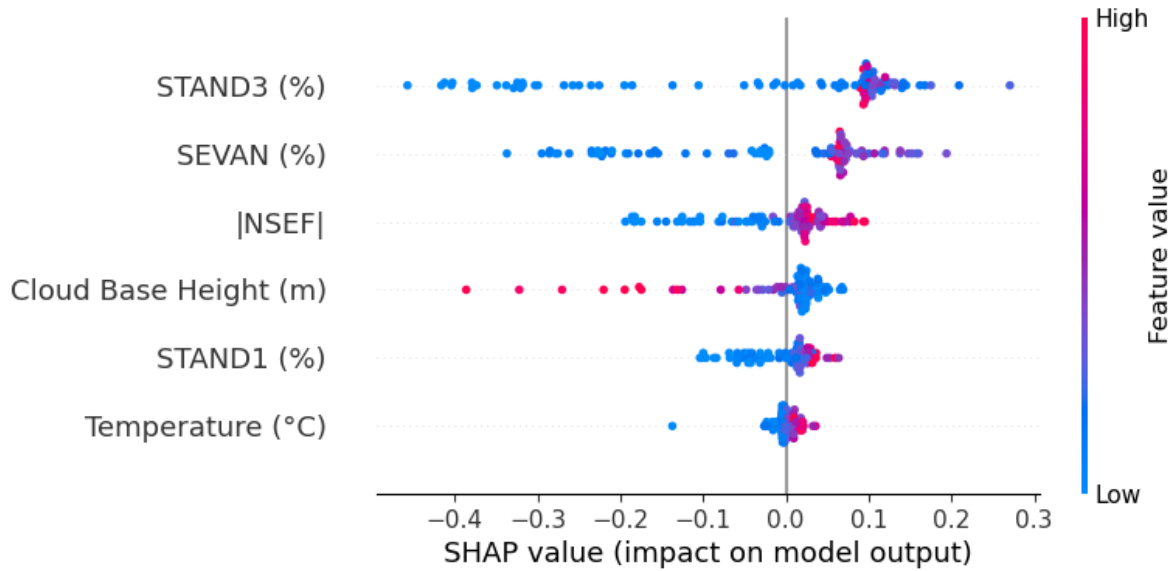

**Figure 8. SHAP summary plot showing the mean absolute SHAP value for each predictor variable.**

Figure 9(a) overlays a LOWESS smoothed curve on the scatter of SHAP values versus the percentage enhancement of STAND 1cm upper scintillator. Below ≈ 4.6 % the smoothed curve remains negative, indicating that small enhancements reduce the likelihood of a TGE relative to the baseline. At ≈4.6 % the curve crosses zero and rises steeply, showing that further increases in count rates of STAND 1cm upper scintillator add substantial positive weight to the TGE prediction. The steep slope beyond the threshold explains why even modest (> 4.6 %) peaks registered by the STAND 1cm scintillator are strong discriminators.

Figure 9(b) presents the SHAP-LOWESS dependence for particle flux percentage enhancement as registered by the upper scintillator of the STAND 3cm detector. The curve crosses zero at approximately 2.45%, showing that relatively modest enhancements in the particle fluxes registered by STAND 3cm detector already contribute positively to TGE classification. Beyond ≈5%, the curve flattens, indicating that further increases yield little additional gain in predictive power.

Figure 9(c) shows that SEVAN percentage enhancements exceeding ≈1.95% result in positive SHAP contributions. The LOWESS curve increases smoothly, without abrupt changes, indicating that even relatively small enhancements in the SEVAN detector strongly influence classification. This is consistent with SEVAN's sensitivity to RREA-produced gamma rays, which contribute measurably even at low flux levels.

Taken together, these results reveal a two-stage behavior. At low to moderate flux enhancements, the SHAP-LOWESS analysis identifies clear thresholds (≈ 2–5 %) beyond which detector responses begin to contribute positively to TGE classification. Once enhancements grow larger, however, the predictive landscape changes qualitatively: for enhancements exceeding ≈10 % in any of the detectors, the classifier achieves perfect separation, with SHAP values uniformly positive and no observed misclassifications. The large percentage increases therefore corresponds to an effectively deterministic signature of TGE activity, highlighting that extreme flux surges are unambiguous indicators of thunderstorm-related particle events.

Figure 10 (a) shows the SHAP dependence on the outside temperature, where the LOWESS curve crosses zero near 0.55°C. Temperatures below this value contribute negatively to TGE likelihood. Figure 10 (b) shows the SHAP dependence on Cloud Base Height. The curve crosses zero at approximately 244m altitude, which is consistent with our previous observations. Figure 10 (c) shows the SHAP dependence on the Near Surface Electric Field values. The SHAP curve becomes strongly positive when the absolute value of the NSEF exceeds ≈7.4 kV/m, indicating that high field variability is a key predictor of TGEs. This aligns very well with the empirical minimum of 5 kV/m used in the manual selection procedure. The statistically derived threshold values are summarized in Table 1.

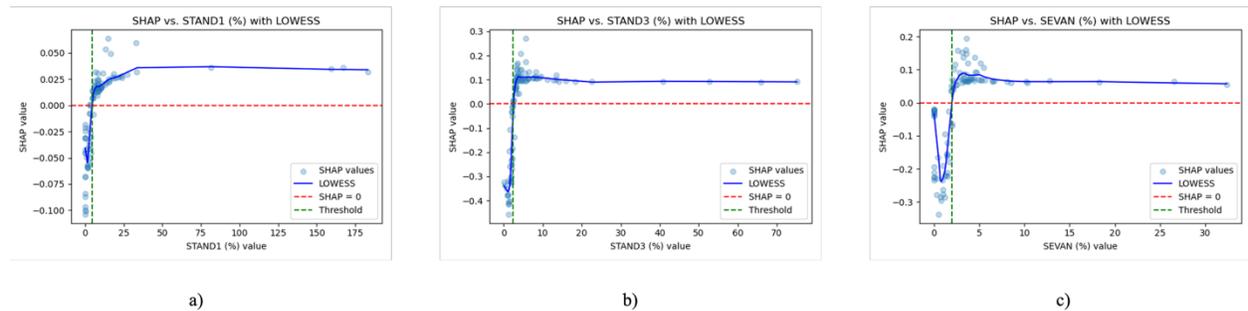

**Figure 9. SHAP dependence plots with LOWESS smoothing lines for a) STAND1 (%), b) STAND3 (%), c) SEVAN (%).**

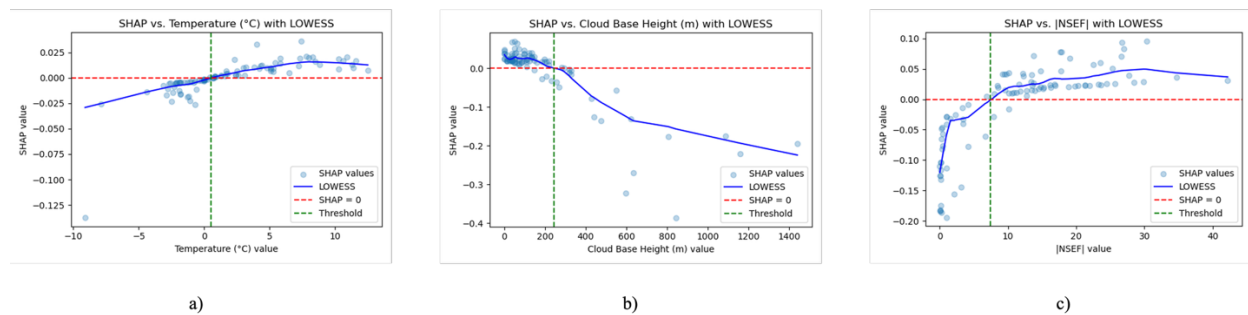

**Figure 10. SHAP dependence plots with LOWESS smoothing lines for a) Temperature (°C), b) Cloud Base Height (m), c) Absolute value of the NSEF (kV/m).**

*Table 1. Summary of approximate threshold values for TGE classification based on TabPFN*

| Parameter | Threshold |
|---|---|
| STAND1 (%) | >= 4.6 |
| STAND3 (%) | >= 2.45 |
| SEVAN (%) | >= 1.95 |
| Temperature (°C) | >= 0.55 |
| Cloud Base Height (m) | <= 244 |
| \|NSEF\| | >= 7.4 |

## 4. Discussion and Conclusions

The SHAP summary (Figure 8) confirms that multi-detector particle enhancement observations, especially the percentage increases recorded by the STAND3 and SEVAN detectors, have the strongest explanatory influence in the TabPFN classifier. Significance-based features ($\sigma$) play a secondary role, indicating that once an enhancement is observed in percentage terms, its statistical significance improves but does not dominate the decision boundary. Environmental variables, such as the near-surface electric field and temperature, further reduce the explanatory gap, showing that even strong particle bursts are unlikely to be classified as TGEs without the proper weather and NSEF conditions. In summer, when high temperatures raise cloud heights, the electric field is also elevated, and TGEs are rare. The SHAP-LOWESS dependence plots (Figure 9, Figure 10) measure these relationships and, importantly, reveal data-driven thresholds (Table 1) that agree well with the rules previously used in manual TGE selection:

- **STAND1 % ≥ 4.6 %, STAND3 % ≥ 2.45 %, SEVAN % ≥ 1.95 %**

confirming that modest enhancements in three independent channels are sufficient for automatic flagging.

Along with the confusion matrix results (Figure 7), these findings show that TGEs occupy a statistically distinct feature space characterized by simultaneous multi-channel enhancements and strong electric field dynamics. The two false negatives highlight that only a small set of borderline cases might be missed by automatic detection, where coincidence-channel review or expert judgment remains important.

The consistency between SHAP-derived thresholds and intuitive selection criteria provides a quantitative validation of fifteen years of empirical practice at ASEC. Importantly, the LOWESS-based approach produces continuous response curves rather than hard cut-offs, enabling adjustable sensitivity should operational priorities shift.

Finally, model generalizability was assessed using the independent dataset of TGEs from 2018 to 2023 (Chilingarian et al., 2022b, 2024b). The TabPFN classifier, which was trained exclusively on 2024 data, correctly identified TGEs with an accuracy of 98.51 %, essentially outperforming its performance on the training-era dataset (94.79 %). The confusion matrix for this evaluation is shown in Figure 11. This robustness highlights the model's capacity to transfer across years and feature sets, further supporting its use in operational TGE identification.

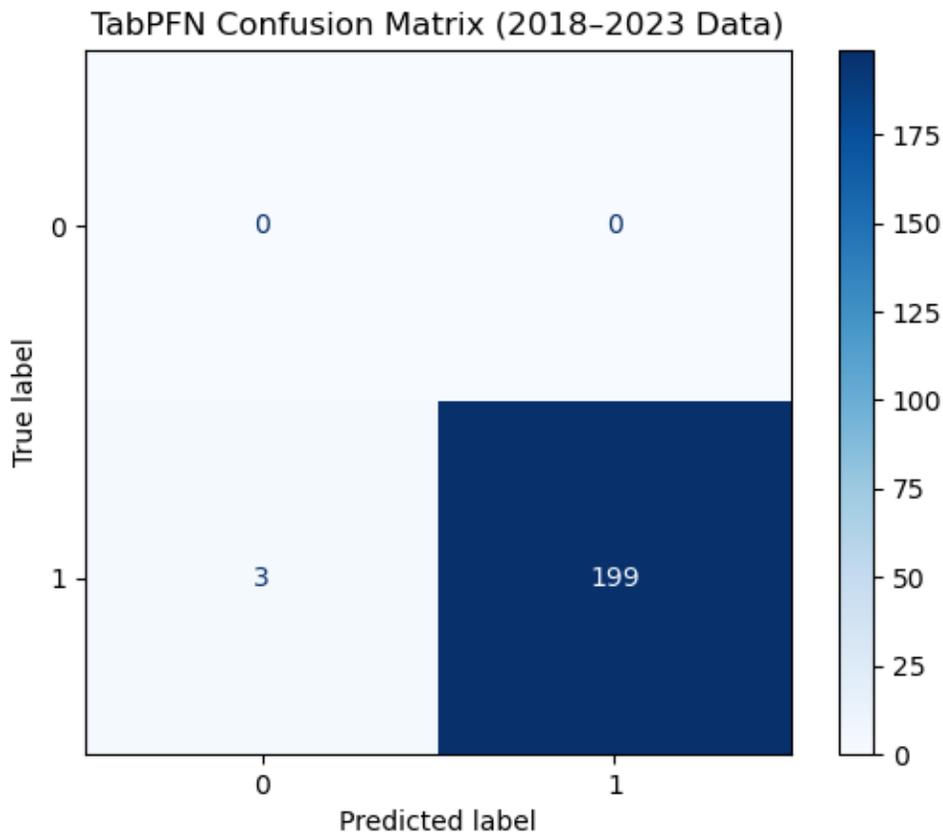

Figure 11. Confusion matrix for the TabPFN classifier on the 2018-2023 TGE set.

This study combines a rigorously curated 2024 TGE dataset with an interpretable TabPFN-SHAP analytical pipeline. Key findings are:

- A high classification accuracy of 94.79% demonstrates that TGEs form a well-defined cluster in feature space.
- Data-driven thresholds extracted via SHAP–LOWESS closely match long-standing intuitive criteria used in 2009-2024, statistically validating used empirical rules.
- Variable importance analysis confirms that observing multi-detector enhancements, combined with environmental variables, are the main predictors of TGEs.

- The framework provides a scalable approach to automate screening, with SHAP thresholds serving as initial filters that identify rare and borderline cases for expert review.

Future work will expand this approach to multi-year, multi-site datasets and explore real-time deployment for operational radiation hazard monitoring and alert issuing.